\documentclass{article}
\usepackage{amsmath}
\usepackage{graphicx}

\input{tcilatex}
\RequirePackage{amsfonts}

\begin{document}

\title{The Interface between Quantum Mechanics and General Relativity}
\author{Raymond Y. Chiao \\
School of Natural Sciences and School of Engineering\\
University of California at Merced}
\date{Lamb Medal Lecture of January 5, 2006\\
(Writeup for the Proceedings of the 36th Winter Colloquium on the Physics of
Quantum Electronics at Snowbird, Utah (version of June 16, 2006 for George
Welch and for the ArXiv))}
\maketitle

\begin{abstract}
The generation, as well as the detection, of gravitational radiation by
means of charged superfluids is considered. \ One example of such a
``charged superfluid'' consists of a pair of Planck-mass-scale, ultracold
``Millikan oil drops,'' each with a single electron on its surface, in which
the oil of the drop is replaced by superfluid helium. When levitated in a
magnetic trap, and subjected to microwave-frequency electromagnetic
radiation, a pair of such ``Millikan oil drops'' separated by a microwave
wavelength can become an efficient quantum transducer between quadrupolar
electromagnetic and gravitational radiation. This leads to the possibility
of a Hertz-like experiment, in which the source of microwave-frequency
gravitational radiation consists of one pair of ``Millikan oil drops''
driven by microwaves, and the receiver of such radiation consists of another
pair of ``Millikan oil drops'' in the far field driven by the gravitational
radiation generated by the first pair. The second pair then back-converts
the gravitational radiation into detectable microwaves. The enormous
enhancement of the conversion efficiency for these quantum transducers over
that for electrons arises from the fact that there exists macroscopic
quantum phase coherence in these charged superfluid systems.
\end{abstract}

\section*{The equivalence principle revisited: Does a falling charge radiate?%
}

Galileo first performed experiments demonstrating that all freely-falling
objects, independent of their mass, accelerate downwards with the same
acceleration \textbf{g} due to Earth's gravity. Later, E\"{o}tv\"{o}s, and
still later, Dicke, performed more sensitive experiments, which showed that
this statement of the equivalence principle was true to extremely high
accuracy, independent of the mass{\normalsize \ }and of the composition of
these objects \cite{Dicke1964}. 
\begin{figure}[ptb]
\label{Galileo}
\par
\begin{center}
\includegraphics[width=3in]{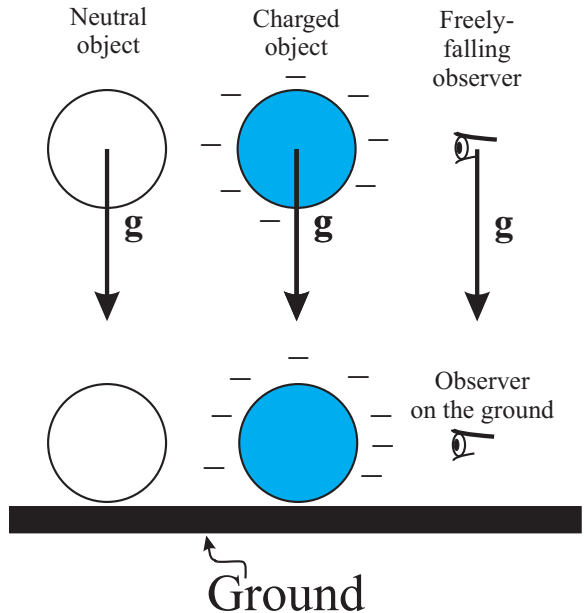}
\end{center}
\caption{Equivalence principle for a neutral and a charged object.}
\end{figure}

One might therefore expect that a neutral object and a charged object, when
simultaneously dropped from the same height, would hit the ground at the
same instant. See Figure 1.

However, a well-known paradox \cite{Boulware1980} now arises when we ask the
following question: Is it the falling charged object, or is it the
stationary charged object at rest on the ground, that radiates
electromagnetic waves?

On the one hand, a freely-falling observer, who is co-moving with the freely
falling neutral and charged objects, sees these two objects as if they were
freely floating in space. The falling charged object would therefore appear
to him not to be accelerating, so that he would conclude that it is not this
charge which radiates. Rather, when he looks downwards at the charged object
which is at rest on the ground, he sees a charge which is accelerating
upwards with an acceleration $-$\textbf{g} towards him. He would therefore
conclude that it is the charged object at rest on the ground, and not the
falling charge, that is radiating electromagnetic radiation.

On the other hand, an observer on the ground would come to the opposite
conclusion. She sees the falling charge accelerating downwards with an
acceleration \textbf{g} towards her, whereas the charged object at rest on
the ground does not appear to her to be undergoing any acceleration. She
would therefore conclude that it is the falling charge which radiates
electromagnetic radiation, and not the charge which is resting on the
ground. Which conclusion is the correct one?

As a first step towards the resolution of this paradox, we note that the
concept of ``radiation'' makes sense only in the far field of moving charged
sources \textit{asymptotically}. We must therefore ask the further question:
What would an observer at infinity see?

Motivated by this further question, let us change the setting for the
formulation of this paradox to that of two nearby objects, one neutral and
one charged, orbiting in free fall around the Earth in the same circular
orbit, as seen by a distant observer. See Figure 2(a). 
\begin{figure}[ptb]
\label{2-earth-orbits}
\par
\begin{center}
\includegraphics[width=3in]{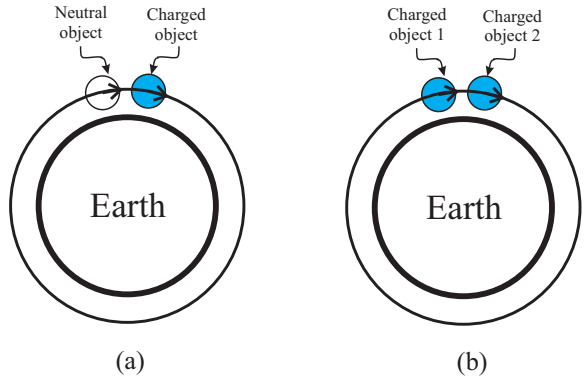}
\end{center}
\caption{(a) Circular orbit around the Earth of a neutral object and a
charged object. (b) Circular orbit of two charged objects.}
\end{figure}

It now becomes clear that the charged object will gradually spiral in
towards the Earth, since it is undergoing constant centripetal acceleration
in uniform circular motion, and will therefore in principle lose energy due
to the emission of electromagnetic radiation at a rate determined by
Larmor's radiation-power formula%
\begin{equation}
P_{EM}=\frac{2}{3}\frac{q^{2}}{c^{3}}a^{2}  \label{EM-Larmor}
\end{equation}
where $P_{EM}$ is the total amount of power emitted in electromagnetic
radiation by the charge $q$ undergoing centripetal acceleration $a$. The
energy escaping to infinity in the form of electromagnetic radiation emitted
by the orbiting charged object must come from its gravitational potential
energy (which is related by the virial theorem to its kinetic energy), and
therefore this object will gradually spiral inwards towards the surface of
the Earth. This kind of decaying orbital motion is the same as that of the
electron in Bohr's planetary model of the hydrogen atom, when the electron's
motion around the proton is considered using only classical concepts. Here
the classical description is clearly a valid one.

Now it is true that the neutral object will also in principle undergo
orbital decay, i.e., it will also gradually spiral inwards towards the
Earth's surface, due to the gradual loss of energy arising from the emission
of gravitational radiation in accordance with the gravitational form of
Larmor's radiation-power formula%
\begin{equation}
P_{GR}^{\prime}=\kappa\frac{2}{3}\frac{Gm^{2}}{c^{3}}a^{2}  \label{GR-Larmor}
\end{equation}
where $\kappa$ is a numerical factor that accounts for the quadrupolar
nature of gravitational radiation, $G$ is Newton's constant, and $m$ is the
mass of the neutral orbiting object, which is undergoing essentially the
same centripetal acceleration $a$ as the charged object \cite{Kappa}. (The
prime on $P_{GR}^{\prime}$\ denotes the incorporation of the factor of $%
\kappa$ into the formula for radiation power.) The decay of orbital motion
due to the emission of gravitational radiation has been observed in the case
of Taylor's binary pulsar PSR 1913+16 \cite{Taylor1994}.

The rate of orbital decay due to the emission of gravitational radiation
will be much smaller than that due to the emission of electromagnetic
radiation, whenever the dimensionless ratio of coupling constants obeys the
inequality%
\begin{equation}
\frac{\kappa Gm^{2}}{q^{2}}<<1.  \label{GM^2<<e^2}
\end{equation}
In such cases, one can neglect the orbital decay due to gravitational
radiation as compared to that due to electromagnetic radiation.

In any case, however, the orbital motion of a charged object will always
decay faster than that of a neutral object with the same mass. An astronaut
would therefore see a differential motion between these two nearby objects.
Even when inside a windowless spacecraft, the astronaut would still be able
to tell the direction of the center of the Earth, by carefully observing the
motion of the charged object relative to the neutral object inside the
spacecraft, since the charged object would be gradually drifting radially
towards the center of the Earth faster than the neutral object. Although
this effect may be extremely small, and may be masked by large systematic
errors, we are discussing here matters of principle. Here the principle of
the conservation of energy demands the existence of this kind of
differential motion.

\section*{Is the equivalence principle violated?}

The above prediction of a differential motion between charged and neutral
objects in Earth's orbit seems at first sight to violate the equivalence
principle, and thus would seem to render invalid the concept of ``geodesic''
in general relativity, which demands that all freely-falling material
objects, independent of their mass and composition (including charge),
traverse the same shortest (geodesic) path in spacetime connecting any two
given spacetime points.

However, it must be kept in mind that the equivalence principle implicitly
assumes that any such material object is to be viewed as a ``vanishingly
small'' test mass, and furthermore implicitly assumes that any charge
associated with this test mass is be to viewed also as a test charge, whose
charge is also ``vanishingly small.'' One is employing here the usual
limiting procedure involving test particles to define the local value of a
classical field, both gravitational and electrical \cite{Test-charge-limit}.

By contrast, a \textit{finitely} charged object experiences a nonvanishing
electromagnetic force due to radiation damping, which is an effectively
viscous kind of force. This implies that a finitely charged object is
undergoing approximately, but not truly exactly, \textit{free} fall. Hence
there is no reason to believe that a finitely charged object would follow a
neutral object along the same geodesic, and the equivalence principle is
therefore not violated.

Charge, at a fundamental level, is a quantum concept. Dirac's
charge-monopole quantization rule shows that the quantization of electrical
charge arises from global, topological, and quantum-mechanical
considerations. The fact that charge is quantized in integer values of the
electron charge $e$, stands in contradiction with the usual limiting
procedure that is used in all classical field theories to define the concept
of ``field,'' in which it is assumed that the test charge (or test mass)
which is used to measure the local value of the field, is a continuous
variable that can be smoothly reduced to zero.

In this classical procedure of taking the test-particle limit, one can
neglect the quantum ``back-action'' of the test particle back onto the
field, because the charge and the mass both smoothly go to zero, and
therefore any back-actions that the test particle might have caused onto the
classical electromagnetic and gravitational fields, must also go smoothly to
zero. However, for a particle with a finite, quantized charge and mass, for
example, for a single electron, this ``no quantum back-action'' assumption
violates the uncertainty principle.

Therefore quantized charged systems are a good place to examine the
conceptual tensions that lie at the interface of quantum mechanics and
general relativity \cite{Chiao2004}. As will be argued below,
single-electron--charged, macroscopically phase-coherent\textit{\ }quantum
fluids are particularly promising systems in which to discover
experimentally new phenomena that might emerge from these conceptual
tensions.

\section*{Two charged objects orbiting the Earth}

Now let us examine the details of the motion of two finitely charged objects
orbiting around the Earth. See Figure 2(b).

For concreteness, imagine that these two charged objects are two Millikan
oil drops with single electrons attached to them, which are nearby to each
other in the same circular orbit. How massive would these oil drops have to
be before the mutual repulsion due to the electrical force between them,
changes to a mutual attraction due to the gravitational force? When they
exceed a certain critical mass, one expects that the drops will drift
towards each other, rather than drifting farther apart. We shall calculate
this critical mass presently.

Now imagine what would happen if a low-frequency gravity wave passes over
these two Millikan oil drops, when this wave propagates at normal incidence
into the plane of the orbit. Such a wave would exert a time-varying tidal
gravitational force, which would alternately stretch and squeeze
sinusoidally in time the space between these objects, when one of the
polarization axes of the gravity wave is chosen to be aligned with respect
to the line connecting the two drops. Therefore the distance between these
charged objects would become an oscillating function of time, according to
the observer at infinity, and this implies the emission of electromagnetic
radiation by these approximately freely-falling objects. Thus this
two-Millikan-oil-drop system would be a kind of transducer, in which
gravitational radiation can be converted into electromagnetic radiation in a
scattering process. \ For weak radiation fields, such a conversion process
would be linear and reciprocal in nature.

However, for very high-frequency gravity waves, it would be possible to
excite a very large number of internal degrees of freedom of the classical
liquid inside a given Millikan oil drop, so that the branching ratio for the
conversion of gravitational wave energy into the electromagnetic wave
channel, as compared to the very large number of possible internal sound and
heat channels, would be extremely small, just as is the case for the
classical Weber bar. For in the reciprocal process, when one attempts to use
a Weber bar as a generator of gravity waves using its fundamental acoustical
mode, the branching ratio for the generation of gravitational radiation
power relative to that of heat generation, has been calculated to be
vanishingly small \cite{Weinberg1972}.

The solution to the problem of the extremely small detection efficiency of
gravitational radiation antennas composed of classical matter, as we shall
argue below, is to freeze out all the internal acoustical and thermal
degrees of freedom of matter at very low temperatures \cite{Freeze-out}, and
to replace the classical matter by macroscopically coherent quantum matter.
For example, instead of the Weber bar, one could use a pair of well
separated, ultracold, levitated singly-charged superfluid helium drops,
where only their center-of-mass degrees of freedom can be excited. There
results a zero-phonon, M\"{o}ssbauer-like motion of an entire superfluid
drop relative to the other drop in response to the application of
high-frequency gravitational or electromagnetic radiation, which can
efficiently generate, as well as detect, gravitational radiation.

In the original M\"{o}ssbauer effect, an excited nucleus of a certain
isotope doped into a crystal can emit a gamma ray, without the usually large
Doppler shift that accompanies the recoil of the emitting nucleus in the
vacuum, because this nucleus is now tightly bound to the lattice. Since the
vibrations of the lattice are quantized into an integer number of phonons,
it is impossible for the system to emit a fraction of a quantum of sound.
There results a large probability that the excited nucleus will emit the
gamma ray in a zero-phonon mode. By the conservation of momentum, the recoil
momentum due to the emission of the radiation must now be taken up by the
center of mass of the entire system. Thus the mass of the recoiling system
is the mass of the entire crystal.

This reduces the recoil Doppler shift by an enormous factor, which is on the
order of the Avogadro's number of atoms present in the entire crystal. The
same enormous factor also reduces the recoil Doppler shift during the
absorption of the gamma ray by an unexcited nucleus of the same isotope,
when this nucleus is also tightly bound to the same lattice. Extremely
narrow gamma-ray resonance-fluorescence lines have therefore been observed
using the same nuclear isotope doped into two separate crystals as emitter
and absorber, one crystal serving as the source, and the other as the
receiver, of the radiation \cite{Mossbauer}.

We shall argue below that a similar M\"{o}ssbauer-like process can occur in
drops of superfluid helium coated with single electrons, when they are
trapped in a strong magnetic field.

\section*{Forces of gravity and electricity between two electrons}

Before going on to the harder problem of electron attachment to superfluid
helium drops, let us first consider the simpler problem of the forces
experienced by two electrons separated by a distance $r$ in the vacuum. Both
the gravitational and the electrical force obey long-range, inverse-square
laws. Newton's law of gravitation states that%
\begin{equation}
\left| F_{G}\right| =\frac{Gm_{e}^{2}}{r^{2}}
\label{Newton's-inverse-square-law}
\end{equation}
where $G$ is Newton's constant and $m_{e}$ is the mass of the electron.
Coulomb's law states that%
\begin{equation}
\left| F_{e}\right| =\frac{e^{2}}{r^{2}}\text{ }  \label{Coulomb's-law}
\end{equation}
where $e$ is the charge of the electron. The electrical force is repulsive,
and the gravitational one attactive.

Taking the ratio of these two forces, one obtains the dimensionless constant%
\begin{equation}
\frac{\left| F_{G}\right| }{\left| F_{e}\right| }=\frac{Gm_{e}^{2}}{e^{2}}%
\approx2.4\times10^{-43}\text{ .}  \label{Gm^2/e^2}
\end{equation}
The gravitational force is extremely small compared to the electrical force,
and is therefore usually omitted in all treatments of quantum physics.

\section*{Gravitational and electromagnetic radiation powers emitted by two
electrons}

The above ratio of the coupling constants $Gm_{e}^{2}/e^{2}$ also is the
ratio of the powers of gravitational to electromagnetic radiation emitted by
two electrons separated by a distance $r$ in the vacuum, when they undergo
an acceleration $a$ relative to each other. Larmor's formula for the power
emitted by a single electron undergoing acceleration $a$ is

\begin{equation}
P_{EM}=\frac{2}{3}\frac{e^{2}}{c^{3}}a^{2}\text{ .}
\end{equation}%
For the case of two electrons undergoing an acceleration $a$ relative to
each other, the radiation is quadrupolar in nature, and the modified Larmor
formula (denoted by the prime) is

\begin{equation}
P_{EM}^{\prime }=\kappa \frac{2}{3}\frac{e^{2}}{c^{3}}a^{2}\text{ ,}
\label{Quad-EM-Larmor}
\end{equation}%
where the prefactor $\kappa $ accounts for the quadrupolar nature of the
emitted radiation. Since the electron carries mass, as well as charge, and
the charge and mass co-move rigidly together, two electrons undergoing an
acceleration $a$ relative to each other will also emit quadrupolar
gravitational radiation according to the formula \cite{Kappa}%
\begin{equation}
P_{GR}^{\prime }=\kappa \frac{2}{3}\frac{Gm_{e}^{2}}{c^{3}}a^{2}\text{ .}
\label{Quad-GR-Larmor}
\end{equation}%
It follows that the ratio of gravitational to electromagnetic radiation
powers emitted by the two-electron system is also given by%
\begin{equation}
\frac{P_{GR}^{\prime }}{P_{EM}^{\prime }}=\frac{Gm_{e}^{2}}{e^{2}}\approx
2.4\times 10^{-43}  \label{Ratio-of-powers}
\end{equation}%
which involves the same ratio of coupling constants as for the ratio of the
forces of gravity to electricity given by Equation (\ref{Gm^2/e^2}). Thus it
would seem at first sight to be hopeless to try and use the two-electron
system as the means for coupling between electromagnetic and gravitational
radiation.

\section*{M{\"o}ssbauer-like response of electron-vortex composites}

However, now consider what would happen if one were to firmly attach an
electron to a vortex at the center of a small circular puddle of a
nanoscale-thick thin film of superfluid helium (i.e., $^{4}$He) adsorbed
onto a cold substrate, which the superfluid does not wet. See Figure 3(a).

Due to the Pauli exclusion principle, the electron forms a nanoscale bubble
inside superfluid helium, which is attracted to the center of the vortex by
the Bernoulli effect. It then forms a bound state with the vortex with the
relatively large binding energy of around 40 K or 3 meV \cite{Donnelly1967}.
In this local minimum-energy configuration, a tightly bound electron-vortex
composite forms at the center of a circular puddle of superfluid, which
possesses a circular boundary since the superfluid does not wet the
substrate. Note the circular symmetry of this system. 
\begin{figure}[ptb]
\label{Mossbauer-like-effect}
\par
\begin{center}
\includegraphics[width=3in]{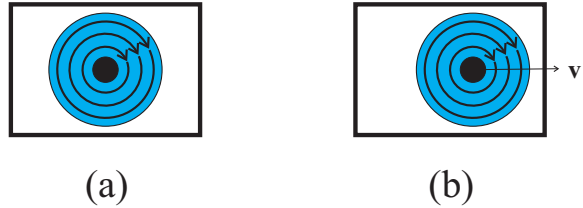}
\end{center}
\caption{(a) An electron (black dot) is tightly bound to a vortex, and forms
an electron-vortex composite system at the center of a circular puddle of a
superfluid helium thin film adsorbed onto a cold, nonwetting substrate
(rectangle). (b) When this system absorbs a microwave photon, the entire
circular puddle recoils in a M{\"{o}}ssbauer fashion.}
\end{figure}

Now imagine what would happen if the electron-vortex system were to absorb a
microwave photon. See Figure 3(b).

In the zero-phonon mode of response, in which no sound waves (nor any other 
\textit{quantized} deformations of the puddle at ultracold temperatures) can
be emitted during the photon absorption process, a given helium atom on the
edge of the puddle cannot cross the circular streamline nearest to the edge.
As a result, this atom is constrained to follow the motion of the vortex
center, along with all the other atoms which make up the entire puddle, in a
M\"{o}ssbauer fashion.

The circular streamlines centered on the electron at the vortex center obey
the quantized-circulation condition given by the Feynman-Onsager rule \cite%
{Donnelly1967}%
\begin{equation}
\oint_{C}\mathbf{v}\cdot d\mathbf{l}=\pm2\pi\frac{\hbar}{m}
\label{Feynman-Onsager}
\end{equation}
where $\mathbf{v}$ is the velocity of the streamline in the vicinity of a
differential line element $d\mathbf{l}$ of a closed curve $C$, $\hbar$ is
Planck's constant, and $m$ is the mass of the helium atom. The physical
meaning of this quantization condition is that there is constructive
interference of each helium atom with itself after one round trip around the
vortex center, such as in any circular path within this kind of matter-wave,
ring-interferometer configuration. The round-trip interference of the helium
atom with itself is similar to that of the photon which occurs in a
ring-laser--gyro configuration.

As a result of being in the zero-phonon mode, the entire electron-vortex
system must recoil as a whole unit in a M\"{o}ssbauer-like response to
external radiation, whenever the system stays adiabatically in its
zero-phonon state, which requires the use of ultralow temperatures \cite%
{Mossbauer}. Thus the mass of the responding system is the mass of the
entire puddle.

Note that the Feynman-Onsager quantization rule is a consequence of the
single-valuedness of the macroscopic wavefunction, i.e., a global quantum
condition that the phase of the macroscopic wavefunction (or ``complex order
parameter'') of the system can only change after one round trip by the
quantized values of $0,$ $\pm2\pi,\pm4\pi,...$\ Furthermore, a vortex is a
topological quantum object with a hole at its center, which possesses a
nonzero winding number of $\pm1$ corresponding to counterclockwise and
clockwise senses of the superflow around the center, respectively. Moreover,
the circulating currents around the vortex center can never stop flowing,
i.e., there exist persistent currents of helium atoms flowing around the
electron trapped at the center of the vortex, that never decay with time.
This is the behavior of a zero-loss, nonviscous charged quantum fluid.

\section*{What's the difference between quantum and classical fluids?}

In light of the above, there are four answers to this question.

(1) A quantum fluid has a ``quantum rigidity'' due to the single-valuedness
of the macroscopic wavefunction, which is absent in classical fluids. London
called this property ``the rigidity of the wavefunction'' in the context of
superconductivity, and Laughlin called\ this property in the context of the
quantum Hall effect ``an incompressible quantum fluid.'' This kind of
rigidity arises because of the quantum adiabatic theorem, which states that
when a quantum many-body system is in its ground state, it will remain
adiabatically in this state in the presence of weak, slowly varying
perturbations, such as those due to weak gravitational or electromagnetic
radiation, provided that there is an energy gap, such as the BCS gap,\ or
the roton gap, or the cyclotron-resonance gap, that separates the ground
state from all possible excited states of the system, so that no transitions
can occur to higher-energy states. Since the search for highly efficient
detectors of gravitational radiation is the search for extremely rigid
matter \cite{Chiao2004}, quantum fluids operating in the M\"{o}ssbauer mode
are good candidates for high-efficiency gravity-wave antennas.

(2) A quantum fluid has a \textquotedblleft quantum
dissipationlessness.\textquotedblright\ The existence of persistent
currents, such as those in the electron-vortex system, is evidence for this
viscosity-free, zero-loss property of a quantum fluid. Hence the generation
of heat in the classical materials used in gravity wave detectors such as
the Weber bar, where heat is an undesirable channel of dissipation of
gravitational wave energy, is automatically closed for such quantum fluids.
Thus in addition to the property of \textquotedblleft quantum
rigidity,\textquotedblright\ the dissipation-free nature of quantum fluids
would allow heat-free motions of superfluid helium drops, for example, in
response to gravitational radiation. This frictionless property of
superfluids would also greatly enhance the conversion efficiency of
gravity-wave detectors based on such fluids, as compared to the extremely
low efficiencies of the highly dissipative Weber bar \cite{Chiao2004}.

(3) The recoil momentum upon the emission or absorption of a microwave
photon by the electron-vortex composite system is taken up by the center of
mass of the whole system in a M\"{o}ssbauer-like effect, which is absent in
a classical fluid. This is yet another aspect of the ``quantum rigidity'' of
the quantum fluid, which does not occur classically.

(4) The entangled state of the electron-vortex system and an emitted
microwave photon generated in the time-reversed version of the
microwave-photon absorption process, would form a bipartite, nonlocal
quantum superposition state which violates Bell's inequalities. Moreover,
due to the interactions among the helium atoms, the quantum many-body system
of the superfluid is automatically in a macroscopically (i.e., massively)
entangled state. The quantum phase coherence of such a macroscopic
superposition state would be quickly destroyed by decoherence in a classical
fluid. However, here decoherence in the superfluid is prevented by the
presence of an energy gap, or more generally, by the presence of a
``scarcity of low-lying states,'' in these ultracold, macroscopically
phase-coherent quantum many-body systems (i.e., ``bosonic quantum fields''),
in what has been called ``gap-protected entanglement'' \cite{Chiao2004}\cite%
{Decoherence}.

\section*{The Planck mass scale}

Let us return to the problem of the ratio of the forces of gravity and
electricity, but now in the context of two well-separated electron-vortex
composites at a distance $r$ from each other.\ See Figure 4. 
\begin{figure}[ptb]
\label{2-electron-vortex-composites}
\par
\begin{center}
\includegraphics[width=3in]{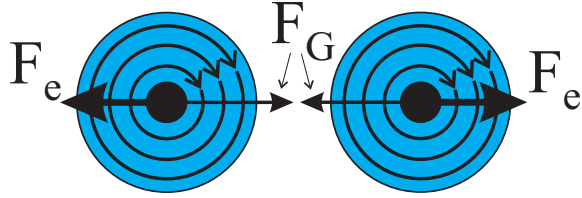}
\end{center}
\caption{Comparison of the attractive gravitational force $F_{G}$ with the
repulsive electrical force $F_{e}$ between two well-separated
electron-vortex composites.}
\end{figure}

Suppose that each circular puddle contains a Planck-mass amount of
superfluid helium, viz.,%
\begin{equation}
m_{\text{Planck}}=\sqrt{\frac{\hbar c}{G}}\approx22\text{ micrograms}
\label{Planck-mass}
\end{equation}
where $\hbar$ is Planck's constant, $c$ is the speed of light, and $G$ is
Newton's constant. Planck's mass sets the characteristic scale at which
quantum mechanics ($\hbar$) impacts relativistic gravity ($c$, $G$). Note
that this mass scale is \textit{mesoscopic} \cite{Kiefer-Weber}, and not
astronomical, in size. This suggests that it may be possible to perform some
novel \textit{nonastronomical}, table-top-scale experiments at the interface
of quantum mechanics and general relativity, which are accessible to the
laboratory.

The ratio of the forces of gravity and electricity between the two
electron-vortex composites now becomes%
\begin{equation}
\frac{\left\vert F_{G}\right\vert }{\left\vert F_{e}\right\vert }=\frac{Gm_{%
\text{Planck}}^{2}}{e^{2}}=\frac{G\left( \hbar c/G\right) }{e^{2}}=\frac{%
\hbar c}{e^{2}}\approx137  \label{137}
\end{equation}
which is 45 orders of magnitude larger than the ratio given earlier by
Equation (\ref{Gm^2/e^2}) for the case of two electrons in the vacuum. Now
the force of gravity is 137 times stronger than the force of electricity, so
that instead of a mutual repulsion between these two charged objects, there
is now a mutual attraction between them. The sign change from mutual
repulsion to mutual attraction between these two electron-vortex composites
occurs at a critical mass $m_{\text{crit}}$ given by%
\begin{equation}
m_{\text{crit}}=\sqrt{\frac{e^{2}}{\hbar c}}m_{\text{Planck}}\approx1.9\text{
micrograms}  \label{m_[crit]}
\end{equation}
whereupon $\left\vert F_{G}\right\vert $ $=\left\vert F_{e}\right\vert $,
and the forces of gravity and electricity balance each other. This is a
strong hint that mesoscopic-scale quantum effects can lead to nonnegligible
couplings between gravity and electromagnetism.

The critical mass $m_{\text{crit}}$ is also the mass at which there occurs a
comparable amount of generation of electromagnetic and gravitational
radiation power upon scattering of radiation from the pair of
electron-vortex composites (or \textquotedblleft Millikan oil
drops,\textquotedblright\ as we shall see below), each member of the pair
with a mass $m_{\text{crit}}$ and a single electron $e$ attached to it. The
ratio of quadrupolar gravitational to the quadrupolar electromagnetic
radiation power ratio is given by%
\begin{equation}
\frac{P_{GR}^{\prime }}{P_{EM}^{\prime }}=\frac{Gm_{\text{crit}}^{2}}{e^{2}}%
=1\text{ ,}  \label{Larmor-power-ratio}
\end{equation}%
where the numerical factors of $\kappa $ in Equations (\ref{EM-Larmor}) and (%
\ref{GR-Larmor}) cancel out, since the charge of the drop co-moves together
with its mass. This implies that the scattered power from these two charged
objects in the gravitational wave channel becomes comparable to that in the
electromagnetic wave channel. However, it should be emphasized that here we
are assuming that the system's charge and mass co-move rigidly together as a
single unit, in accordance with the M\"{o}ssbauer-like mode of response to
radiation fields. This is purely quantum effect based on the quantum
adiabatic theorem's prediction that the system will remain adiabatically in
its nondegenerate ground state.

\section*{Simplification to ``Millikan oil drops''}

From now on, we shall use the term ``Millikan oil drop'' with quotation
marks (or, drop, without quotation marks), as the abbreviated nomenclature
for ``Planck-mass-scale superfluid-helium drop with a single electron firmly
attached to its surface, which exhibits a M\"{o}ssbauer-like response to the
application of high-frequency radiation fields.'' By going from the 2D thin
superfluid-helium film geometry of the electron-vortex composite to that a
3D superfluid-helium drop, we avoid experimental complications arising from
the choice of wetting versus non-wetting substrates, and all other such
substrate-related physics.

Although for simplicity we shall first consider ``Millikan oil drops'' with
only a single electron attached to each drop, there is no reason not to
consider the case also where many electrons are attached to each drop, and
where a quantum Hall fluid forms on the surface of the drop in the presence
of a strong magnetic field, as long as the charge-to-mass ratio of the drop
is kept fixed so that the condition%
\begin{equation}
\frac{P_{GR}^{\prime}}{P_{EM}^{\prime}}=1  \label{Equality-of-powers}
\end{equation}
is still satisfied. The quantum many-body system of the many-electron drop
at ultra-low temperatures will go into its ground state, and can still
possess a macroscopic amount of gravitational mass.

The helium atom is diamagnetic, and liquid helium drops have successfully
been magnetically levitated in an anti-Helmholtz magnetic trapping
configuration \cite{Weilert1996}. Due to its surface tension, the surface of
a freely suspended, ultracold superfluid drop is atomically perfect. When an
electron is approaches a drop, the formation of an image charge inside the
dielectric sphere of the drop causes the electron to be attracted by the
Coulomb force to its own image. However, the Pauli exclusion principle
prevents the electron from entering the drop. As a result, the electron is
bound\ to the surface of the drop in a hydrogenic ground state.
Experimentally, the binding energy of the electron to the surface of liquid
helium has been measured using millimeter-wave spectroscopy to be 8 Kelvin 
\cite{Grimes}, which is quite large compared to the milli-Kelvin temperature
scales for the proposed experiment. Hence the electron is tightly bound to
the surface of the drop.

Such a \textquotedblleft Millikan oil drop\textquotedblright\ is just as
much a macroscopically phase-coherent quantum object as is the
electron-vortex composite discussed earlier. In its ground state, which
possesses a single, coherent quantum mechanical phase throughout the
interior of the superfluid, the drop possesses a zero circulation quantum
number (i.e., contains no quantum vortices), with one unit (or an integer
multiple) of the charge quantum number. As a result of the drop being at
ultra-low temperatures, all degrees of freedom other than the center-of-mass
degrees of freedom are frozen out, so that there results a zero-phonon M\"{o}%
ssbauer-like effect, in which the entire mass of the drop moves rigidly as a
single unit in response to radiation fields. Also, since it remains
adiabatically in the ground state during weak, but possibly arbitrary,
perturbations due to these radiation fields, the \textquotedblleft Millikan
oil drop,\textquotedblright\ like the electron-vortex composite,\ possesses
a quantum rigidity and a quantum dissipationlessness that are the two most
important quantum properties for achieving a high conversion efficiency for
gravity-wave antennas.

Note that a pair of spatially separated ``Millikan oil drops'' have the
correct quadrupolar symmetry in order to couple to gravitational radiation,
as well as to quadrupolar electromagnetic radiation. When they are separated
by a distance on the order of a wavelength, they become an efficient
quadrupolar antenna for generating, as well as detecting, gravitational
radiation.

\section*{A pair of ``Millikan oil drops'' as a transducer}

Let us now place a pair of \textquotedblleft Millikan oil
drops\textquotedblright\ separated by approximately a microwave wavelength
inside a black box, which represents a quantum transducer that can convert
gravitational (GR) waves into electromagnetic (EM) waves, as indicated
schematically in Figure 5(a). This kind of transducer action is similar to
that discussed earlier for a low-frequency gravity wave passing over a pair
of charged, freely falling objects orbiting the Earth indicated in Figure
2(b). It should again be stressed that these finitely charged, and
approximately, but not truly exactly, freely falling objects in fact do
radiate electromagnetic waves, since these waves are observable by an
observer at infinity, and that the emission of this radiation does not lead
to a violation of the equivalence principle, as was discussed earlier. 
\begin{figure}[ptb]
\begin{center}
\label{Transducer} \includegraphics[width=4in]{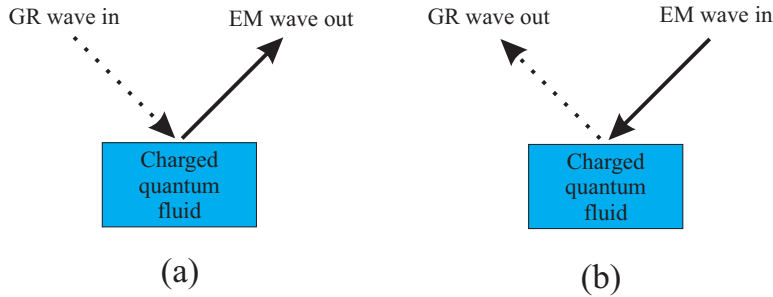}
\end{center}
\caption{(a) A charged quantum fluid acts as a transducer that converts
gravity waves into electromagnetic waves. (b) The reciprocal transducer
action that converts electromagnetic waves into gravity waves.}
\end{figure}

By time-reversal symmetry, the reciprocal process (b), as indicated in
Figure 5(b), in which a charged quantum fluid such as another pair of
``Millikan oil drops,'' converts an electromagnetic wave into a
gravitational wave, must also occur with the same efficiency as the forward
process (a) of Figure 5(a). The time-reversed (or ``back-action'') process
(b) is important because it allows the \textit{generation} of gravitational
radiation, and can therefore become a practical \textit{source} of such
radiation.

\section*{Hertz-like experiment}

This raises the possibility of performing a Hertz-like experiment, in which
process (b) becomes the source, and its reciprocal process (a) becomes the
receiver, of gravity waves, as indicated in Figure 6. 
\begin{figure}[ptb]
\label{Hertz-like}
\par
\begin{center}
\includegraphics[width=3.5in]{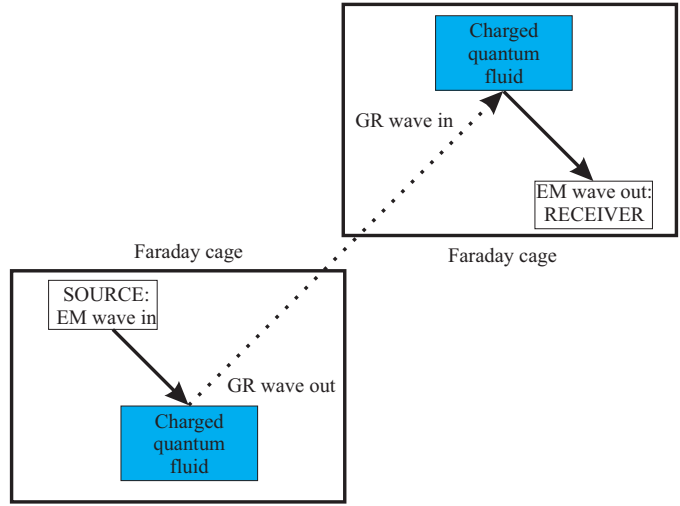}
\end{center}
\caption{A Hertz-like experiment, in which a quantum transducer converts
electromagnetic (EM) waves into gravity (GR) waves, and a second quantum
transducer in the far field of the first back-converts gravity (GR) waves
into detectable electromagnetic (EM) waves.}
\end{figure}

Room-temperature Faraday cages, indicated by rectangles in Figure 6, prevent
the transmission of electromagnetic waves, so that only gravitational waves,
which can easily pass through all classical matter such as the normal (i.e.,
dissipative)\ metals of which standard, room-temperature Faraday cages are
composed, are transmitted between the two halves of the apparatus that serve
as the source and the receiver, respectively. Such an experiment would be
practical to perform using standard microwave sources and receivers, if the
scattering cross-sections and the transducer conversion efficiencies of the
two charged quantum fluids are not too small.

An experiment using YBCO, which is a superconductor at liquid nitrogen
temperatures, as the material for the two charged quantum-fluid transducers
in the Hertz-like experiment, has been performed at 12 GHz \cite{Chiao2004}.
The conversion efficiency of each YBCO transducer in the two-transducer
system, assuming that the two transducers are identical, has been measured
to be less than 15 parts per million (probably due to the high microwave
losses of YBCO, as compared to the extremely low characteristic impedance of
free space for gravity waves, $Z_{G}=16\pi G/c=1.1\times10^{-17}$ SI units 
\cite{Chiao2004}).

\section*{M{\"{o}}ssbauer-like response of \textquotedblleft Millikan oil
drops\textquotedblright\ in a magnetic trap to radiation fields}

As a more practical realization of a quantum transducer using a charged
quantum fluid, let us consider a pair of levitated \textquotedblleft
Millikan oil drops\textquotedblright\ in a magnetic trap, where the drops
are separated by a distance on the order of a microwave wavelength, which is
chosen so as to satisfy the impedance-matching condition for a good
quadrupolar microwave antenna. See Figure 7. 
\begin{figure}[ptb]
\label{2-oil-drops-in-trap}
\par
\begin{center}
\includegraphics[width=3in]{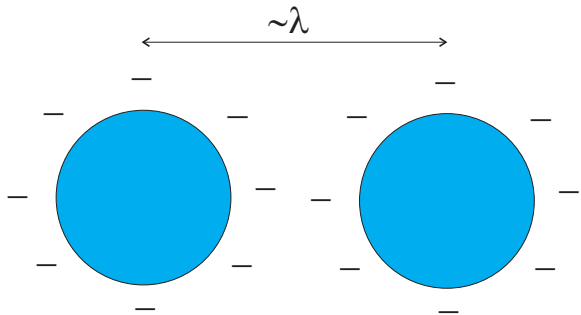}
\end{center}
\caption{Two levitated \textquotedblleft Millikan oil
drops\textquotedblright\ in a magnetic trap.}
\end{figure}

Now let a beam of electromagnetic waves in the Hermite-Gaussian TEM$_{11}$
mode \cite{Yariv1967}, which has a quadrupolar transverse field pattern that
has a substantial overlap with that of a gravitational plane wave, impinge
at a 45$^{\circ}$ angle with respect to the line joining these two charged
objects, as indicated in Figure 6. As a result of being thus irradiated, the
pair of \textquotedblleft Millikan oil drops\textquotedblright\ will be
driven into motion in an anti-phased manner, so that the distance between
them will oscillate sinusoidally with time, according to an observer at
infinity \cite{Gremlin}. Thus the simple harmonic motion of the two drops
relative to one another produces a time-varying mass quadrupole moment at
the same frequency as that of the driving electromagnetic wave. This
oscillatory motion will in turn scatter (in a linear scattering process) the
incident electromagnetic wave into gravitational and electromagnetic
scattering channels with comparable powers, provided that the ratio of
quadrupolar Larmor radiation powers given by Equation (\ref%
{Larmor-power-ratio}) is of the order of unity, which will be case when the
mass of both drops is on the order of the critical mass $m_{\text{crit}}$
for the case of single electrons attached to each drop. The reciprocal
process should also have a power ratio of the order of unity.

We will now discuss the M{\"{o}}ssbauer-like response of \textquotedblleft
Millikan oil drops.\textquotedblright\ Imagine what would happen if one were
replace an electron in the vacuum with a single electron which is firmly
attached to the surface of a drop of superfluid helium ($^{4}$He) in the
presence of a strong magnetic field and at ultralow temperatures, so that
the system of the electron and the superfluid, considered as a single
quantum entity, would form a macroscopic quantum ground state. Such a
quantum system can possess a sizeable gravitational mass. For the case of
many electrons attached to a massive drop, where a quantum Hall fluid forms
on the surface of the drop in the presence of a strong magnetic field, there
results a nondegenerate, Laughlin-like ground state.

In the presence of Tesla-scale magnetic fields, an electron is effectively
prevented from moving at right angles to the local magnetic field line
around which it is executing tight cyclotron orbits. The result is that the
surface of the drop, to which the electron is tightly bound, cannot undergo
liquid-drop deformations, such as the oscillations between the prolate and
oblate spheroidal configurations of the drop which would occur at low
frequencies in the absence of the magnetic field. After the drop has been
placed into Tesla-scale magnetic fields at milli-Kelvin operating
temperatures, both the single- and many-electron drop systems will be
effectively frozen into the ground state, since the characteristic energy
scale for electron cyclotron motion in Tesla-scale fields is on the order of
Kelvins. Due to the tight coupling of the electron(s) to the surface of the
drop, this would freeze out all shape deformations of the superfluid drop.

Since all internal degrees of freedom of the drop, such as its microwave
phonon excitations, will also be frozen out at sufficiently low
temperatures, the charge and the entire mass of the \textquotedblleft
Millikan oil drop\textquotedblright\ should co-move rigidly together as a
single unit, in a M\"{o}ssbauer-like response to applied radiation fields.
This is a result of the elimination of all internal degrees of freedom by
the Boltzmann factor at sufficiently low temperatures, so that the system
stays in its ground state, and only the external degrees of freedom of the
drop, consisting only of its center-of-mass motions, remain.

The criterion for this \textquotedblleft zero-phonon,\textquotedblright\ or M%
\"{o}ssbauer-like, mode of response of the electron-drop system is that the
temperature of the system is sufficiently low, so that the probability for
the entire system to remain in its nondegenerate ground state without even a
single quantum of excitation of any of its internal degrees of freedom being
excited, is very high, i.e.,%
\begin{equation}
\text{Prob. of zero internal excitation}\approx 1-\exp \left( -\frac{E_{%
\text{gap}}}{k_{B}T}\right) \rightarrow 1\text{ as }\frac{k_{B}T}{E_{\text{%
gap}}}\rightarrow 0,  \label{Prob(no excitation)}
\end{equation}%
where $E_{\text{gap}}$ is the energy gap separating the nondegenerate ground
state from the lowest permissible excited states, $k_{B}$ is Boltzmann's
constant, and $T$ is the temperature of the system. Then the quantum
adiabatic theorem ensures that the system will stay adiabatically in the
nondegenerate ground state of this quantum many-body system during
perturbations, such as those due to weak, externally applied radiation
fields. By the principle of momentum conservation, since there are no
internal excitations to take up the radiative momentum, the center of mass
of the entire system must undergo recoil in the emission and absorption of
radiation. Thus the mass involved in the response to radiation fields is the
mass of the whole system.

For the case of a single electron (or many electrons in the case of the
quantum Hall fluid)\ in a strong magnetic field, the typical energy gap is
given by%
\begin{equation}
E_{\text{gap}}=\hbar\omega_{\text{cyclotron}}=\frac{\hbar eB}{mc}>>k_{B}T%
\text{ ,}  \label{Cyclotron-gap}
\end{equation}
an inequality which is valid for the Tesla-scale fields and milli-Kelvin
temperatures being considered here.

\section*{Estimate of the scattering cross-section}

Let $d\sigma _{a\rightarrow \beta }$ be the differential cross-section for
the scattering of a mode $a$ of radiation of an incident gravitational wave
to a mode $\beta $ of a scattered electromagnetic wave by a pair of
\textquotedblleft Millikan oil drops.\textquotedblright\ (We shall denote GR
waves by Roman-letter subscripts, and EM waves by Greek-letter subscripts.)
Then, by time-reversal symmetry%
\begin{equation}
d\sigma _{a\rightarrow \beta }=d\sigma _{\beta \rightarrow a}\text{ .}
\end{equation}%
Since electromagnetic and weak gravitational fields both formally obey
Maxwell's equations \cite{Wald} (apart from a difference in the signs of the
source density and the source current density), and since these fields obey
the same boundary conditions, the solutions for the modes for the two kinds
of scattered radiation fields must also have the same mathematical form. Let 
$a$ and $\alpha $ be a pair of corresponding solutions, and $b$ and $\beta $
be a different pair of corresponding solutions to Maxwell's equations for GR
and EM modes, respectively. For example, $a$ and $\alpha $ could represent
incoming plane waves which copropagate in the same direction, and $b$ and $%
\beta $ scattered, outgoing plane waves which copropagate together in a
different direction. Then for the case of a pair of critical-mass drops with
single-electron attachment, there is an equal conversion into the two types
of scattered radiation fields in accordance with Equation (\ref%
{Larmor-power-ratio}), and therefore%
\begin{equation}
d\sigma _{a\rightarrow b}=d\sigma _{a\rightarrow \beta }\text{ ,}
\end{equation}%
where $b$ and $\beta $ are corresponding modes of the two kinds of scattered
radiations.

By the same line of reasoning, for this pair of critical-mass drops%
\begin{equation}
d\sigma _{b\rightarrow a}=d\sigma _{\beta \rightarrow a}=d\sigma _{\beta
\rightarrow \alpha }\text{ .}
\end{equation}%
It therefore follows from the principle of reciprocity (i.e. time-reversal
symmetry) that%
\begin{equation}
d\sigma _{a\rightarrow b}=d\sigma _{\alpha \rightarrow \beta }.
\end{equation}

In order to estimate the size of the total cross-section, it is easier to
consider first the case of electromagnetic scattering, such as the
scattering of microwaves from two critical-mass drops, with radii $R$ and a
separation $r$ on the order of a microwave wavelength. Let the electrons on
the \textquotedblleft Millikan oil drops\textquotedblright\ be in a quantum
Hall plateau state, which we know is that of a perfectly dissipationless
quantum fluid, like that of a superconductor. Furthermore, we know that the
nondegenerate Laughlin ground state is that of a perfectly rigid,
incompressible quantum fluid \cite{Laughlin}. The two drops thus behave like
perfectly conducting, shiny, mirrorlike spheres, which scatter light in a
manner similar to that of perfectly elastic hard-sphere scattering in
idealized billiards. The total cross section for the scattering of
electromagnetic radiation from a pair of drops is therefore given
approximately by the geometric cross-sectional areas of two hard spheres%
\begin{equation}
\sigma _{\alpha \rightarrow \text{all }\beta }=\int d\sigma _{\alpha
\rightarrow \beta }\simeq \text{Order of }2\pi R^{2}
\label{Geometric-X-section}
\end{equation}%
where $R$ is the hard-sphere radius of a drop.

However, if, as one might expect on the basis of classical intuitions, that
the total cross-section $\sigma _{a\rightarrow \text{all }b}$ for the
scattering of gravitational waves from the two-drop system would be
extremely small, like that of all classical matter such as the Weber bar,
then by reciprocity, the total cross-section $\sigma _{\alpha \rightarrow 
\text{all }\beta }$ for the scattering of electromagnetic waves from the
two-drop system must also be extremely small. This would lead to a
contradiction with the hard-sphere cross section given by Equation (\ref%
{Geometric-X-section}), so these intuitions must be incorrect.

From the reciprocity principle and from the important properties of quantum
rigidity and quantum dissipationlessness of these drops, one therefore
concludes that for two critical-mass \textquotedblleft Millikan oil
drops,\textquotedblright\ it must be the case that%
\begin{equation}
\sigma _{a\rightarrow \text{all }b}=\sigma _{\alpha \rightarrow \text{all }%
\beta }\simeq \text{Order of }2\pi R^{2}\text{ .}
\end{equation}%
\qquad

\section*{Signal-to-noise considerations}

The signal-to-noise ratio expected for the Hertz-like experiment depends on
the current status of microwave source and receiver technologies. Based on
the experience gained from the experiment done on YBCO using existing
off-the-shelf microwave components \cite{Chiao2004}, we expect that we would
need geometric-sized cross-sections and a minimum conversion efficiency on
the order of a few parts per million per transducer, in order to detect a
signal.

It should be stressed that in the Hertz-like experiment, one is not trying
to detect the \textit{strain} of space (which may be extremely small), but
rather the \textit{power} that is being transferred by radiation from one
quantum transducer to the other. The overall signal-to-noise ratio depends
on the initial microwave power, the scattering cross-section, the conversion
efficiency of the quantum transducers, and the noise temperature of the
microwave receiver (i.e., its first-stage\ amplifier).

Microwave low-noise amplifiers can possess noise temperatures that are
comparable to room temperature (or even better, such as in the case of
liquid-helium cooled paramps used in radio astronomy). The minimum power $%
P_{\min}$ detectable in an integration time $\tau$ is given by%
\begin{equation}
P_{\min}=\frac{k_{B}T_{\text{noise}}\Delta\nu}{\sqrt{\tau\Delta\nu}}
\end{equation}
where $k_{B}$ is Boltzmann's constant, $T_{\text{noise}}$ is the noise
temperature of the first stage microwave amplifier, and $\Delta\nu$ is its
bandwidth. Assuming an integration time of one second, and a bandwidth of 1
GHz, and a noise temperature $T_{\text{noise}}=300$ K, one gets $%
P_{\min}(\tau=$1 sec$)=1.3\times10^{-25}$ Watts.

\section*{Why such an enormous enhancement?}

Why is there such an enormous enhancement of over 40 orders of magnitude in
the quantum transducer conversion efficiency predicted by Equation (\ref%
{Larmor-power-ratio}) for two \textquotedblleft Millkan oil
drops\textquotedblright\ over that for two electrons in the vacuum separated
by the same distance, predicted by Equation (\ref{Ratio-of-powers})?

The answer is that the macroscopic quantum phase coherence of superfluid
helium allows an enormous number of atoms in the superfluid to all move
together coherently in unison in response to gravitational radiation, so
that there exists an enormous enhancement of the oscillating mass quadrupole
moment of the two drops by a factor of $N_{atom}$, the number of atoms
participating in the time-varying superposition of the center-of-mass
momentum eigenstates of these drops induced by the radiation. There is a
corresponding enhancement in the amount of gravitational radiation power
that is emitted by a pair of \textquotedblleft Millikan oil
drops\textquotedblright\ over that emitted by a pair of bare electrons
separated by the same distance in the vacuum, by a factor of $N_{atom}^{2}$.
In the case of the Planck mass, $N_{atom}\sim 10^{18}$ helium atoms, and in
the case of the critical mass, $N_{atom}\sim 10^{17}$ helium atoms. At a
fundamental level, this enormous enhancement originates from the
superposition principle of quantum mechanics.

Here I am assuming that there does not exist any appreciable decoherence of
quantum superposition states which contain a sufficiently large amount of
gravitational mass so that the superposition principle of quantum mechanics
breaks down. It has been suggested that stochastic backgrounds of gravity
waves from the Big Bang acting on quantum superpositions at the Planck mass
scale may indeed cause such a decoherence \cite{Reynaud2006}. The Hertz-like
experiment, if properly performed, may be a test of the validity of the
superposition principle of quantum mechanics for Planck-mass objects such as
\textquotedblleft Millikan oil drops.\textquotedblright\ I hope to be able
to perform the Hertz-like experiment with my colleagues at Merced.

\section*{Acknowledgments}

My appreciation goes to Marlan Scully for awarding me the Willis Lamb medal.
I would also like to thank Gene Commins, Marc Feldman, Giorgio Frossati,
John Garrison, Dan Gauthier, {\O }yvind Gr{\o }n, Dave Kelley, Wolfgang
Ketterle, Manfred Kleber, Jon Magne Leinaas, Robert Littlejohn, Peter
Milonni, Kevin Mitchell, Richard Packard, Bill Phillips, Paul Richards, Ray
Sachs, Wolfgang Schleich, Achilles Speliotopoulos, Cliff Taubes, Peter
Toennies, Dave Wineland, and Roland Winston for helpful discussions.\ I
would also like to thank all my former students, especially Ivan Deutsch,
Paul Kwiat, and Aephraim Steinberg, for organizing the recent
\textquotedblleft Chiaofest,\textquotedblright\ and to thank all my
students, collaborators and colleagues, especially Charles Townes, for
honoring me by their presence at this celebration in Snowbird.


\begin{thebibliography}{99}
\bibitem{Freeze-out} By the ``freezing-out'' of a degree of freedom, we mean
here that the Boltzmann factor becomes exponentially small for temperatures
less than the lowest possible excitation energy associated with this degree
of freedom. For example, the vibrational degree of freedom of molecular
hydrogen is frozen out exponentially below 6000 K, so that the molecule
behaves effectively like a perfectly rigid dumbbell below this temperature,
as is observed in the behavior of the molar specific heat as a function of
temperature.
\bibitem{Decoherence} Two kinds of ``decoherence'' should be distinguished:
the first is the decoherence which is prevented from occurring by the
``gap-protected entanglement'' of a superconductor or superfluid, in which
the phase of the macroscopic order parameter becomes well defined below the
critical temperature or lambda point; the second is the decoherence of a
macroscopic superposition state, such as that of two opposite flux states of
a superconducting ring with a Josephson junction in it, i.e., a
superconducting ``qubit.'' The first kind of quantum superposition is robust
against environmental decoherence because of the gap, but the second kind of
superposition is not. I thank Marc Feldman for pointing this out to me.
\end{thebibliography}
\end{document}